\documentclass[nofootinbib,prl,amsmath,amssymb,twocolumn,floatfix]{revtex4-2}

\usepackage{graphicx}
\usepackage{dcolumn}
\usepackage{bm}
\usepackage{siunitx}
\usepackage{todonotes}
\usepackage[hidelinks]{hyperref}

\begin{document}

\preprint{APS/123-QED}



\title{A Two-Scale Effective Model for Defect-Induced Localization Transitions in Non-Hermitian Systems}

\author{B. Davies$^{1,2}$, S. Barandun$^{3}$, E.~O. Hiltunen$^{4}$, R.~V. Craster$^{1,2,5,6}$ and H. Ammari$^{3}$}
\affiliation{$^1$Department of Mathematics, Imperial College London, London SW7 2AZ, United Kingdom \\
$^2$Centre for Plasmonics and Metamaterials, Imperial College London, London SW7 2AZ, United Kingdom \\
$^3$Department of Mathematics, ETH Zurich, Zurich 8092, Switzerland \\
$^4$Department of Mathematics, Yale University, New Haven, CT 06511, USA \\
$^5$Department of Mechanical Engineering, Imperial College London, London SW7 2AZ, United Kingdom \\
$^6$UMI 2004 Abraham de Moivre-CNRS, Imperial College London, London SW7 2AZ, United Kingdom}

\date{\today}

\begin{abstract}
We illuminate the fundamental mechanism responsible for the transition between the non-Hermitian skin effect and defect-induced localization in the bulk. We study a Hamiltonian with non-reciprocal couplings that exhibits the skin effect (the localization of all eigenvectors at one edge) and add an on-site defect in the center. Using a two-scale asymptotic method, we characterize the long-scale growth and decay of the eigenvectors and derive a simple and intuitive effective model for the transition that occurs when the defect is sufficiently large that one of the modes is localized at the defect site, rather than at the edge of the system.
\end{abstract}


\maketitle

\textit{Introduction}.---Understanding and harnessing systems that are non-Hermitian but have real spectra represents one of the most exciting and active frontiers in the physical sciences \cite{okuma2023non, ashida2020non, bergholtz2021exceptional, bender1998real, bender2007making}. For example, lattices with non-reciprocal couplings have been studied at length for their ability to support spectra composed exclusively of eigenmodes that are exponentially localized at one edge of the system \cite{yao2018edge, zhang2022review, lin2023topological, li2020critical, song2019nonHerm, zhang2021acoustic, maddi2024exact, scheibner2020non, yao2018edge, lee2019anatomy, kawabata2020higher, zhang2021observation, zhang2022universal, ammari2023skin3d}. This phenomenon, known as the \emph{non-Hermitian skin effect}, has deep implications for localization and transport properties and has been realised in settings including quantum systems \cite{song2019nonHerm}, acoustic metamaterials \cite{zhang2021acoustic, maddi2024exact}, elastic media \cite{scheibner2020non}, dimerized systems \cite{yao2018edge, lee2019anatomy} and multi-dimensional models \cite{kawabata2020higher, zhang2021observation, zhang2022universal, ammari2023skin3d}.

The non-Hermitian skin effect is a very strong localization phenomenon and there is significant interest in understanding how it behaves in the presence of imperfections and disorder \cite{hatano1996localization, hatano1998nonHermitian, brouwer1997theory, goldsheid2018real, longhi2019probing, borgnia2020non, okuma2020topological, alvarez2018non, ammari2023stability}. The effect is known to persist in the presence of small disorder; however, when sufficiently large defects are added, Anderson-type effects can cause eigenvectors to be localized within the bulk of the system (away from the edge). The most common setting in which to study these effects is the non-Hermitian Anderson model, which is typically a nearest-neighbour Hamiltonian with a random potential (independent on each site) and non-reciprocal coupling strengths \cite{hatano1996localization, hatano1998nonHermitian}. Many studies have considered this model's spectral properties in the presence of periodic boundary conditions, to understand its topological features and how its complex spectrum behaves in the presence of disorder \cite{hatano1996localization, hatano1998nonHermitian, brouwer1997theory, goldsheid2018real, longhi2019probing, borgnia2020non, okuma2020topological}. However, one of the most exciting features of non-Hermitian systems is that altering the boundary conditions can drastically affect their bulk properties \cite{okuma2020topological}. Indeed, in a finite-sized model with open or fixed boundary conditions at the edges, the skin effect can occur \cite{okuma2023non, yao2018edge, zhang2022review, lin2023topological, li2020critical, song2019nonHerm, zhang2021acoustic, maddi2024exact, scheibner2020non, yao2018edge, lee2019anatomy, kawabata2020higher, zhang2021observation, zhang2022universal, ammari2023skin3d}. In this case, when imperfections are introduced there is the additional challenging question of predicting where eigenvectors will be localized since the site where a given eigenvector is localized is the result of competition between defect-induced localization in the bulk and localization due to the skin effect at the edge. This Letter develops an effective model to illuminate this competition and the resulting transition between the two regimes.

The defect-induced localization transition in non-Hermitian systems is fundamentally different from classical Hermitian analogues \cite{makwana2013localised, craster2023asymptotic}. In Hermitian settings, if localization within the bulk occurs, it will typically do so for any arbitrarily small defect size. In our non-Hermitian setting, however, the localizing strength of the skin effect means that small defects do not induce localization within the bulk. Instead, there is some non-zero critical defect size at which the defect is sufficiently large to induce localization. This transition can be predicted from the topological origins of the non-Hermitian skin effect \cite{borgnia2020non, okuma2020topological, ammari2023mathematical}. Due to the theory of Toeplitz operators, eigenvectors corresponding to eigenvalues falling within a specified region of the complex plane (specified by the winding number of the associated symbol) must be localized at the edge \cite{bottcher2012introduction, trefethen2005Spectra}. The boundary of this region is where the localization transition occurs as eigenmodes escaping it (due to the influence of defects) can be localized within the bulk. This can be used to quantify the robustness of the skin effect \cite{alvarez2018non, ammari2023stability} and to characterize the critical defect size at which an eigenvalue crosses the boundary and a localization transition occurs.

In this Letter, the localization transition due to defects in a system exhibiting the non-Hermitian skin effect is revealed. We develop a two-scale effective model for the localization transition that occurs when sufficiently large defects are added to a lattice with non-reciprocal couplings. This characterizes the competition between localization at the edge due to the non-Hermitian skin effect and defect-induced localization in the bulk. Our approach to characterizing the effective properties of the system near to this localization transition is a variant of the two-scale asymptotic methods known variously as $\mathbf{k}\cdot\mathbf{p}$ theory \cite{voonwillatzen2009kp, janssen2016precise} or high-frequency homogenization (HFH) \cite{craster2010high}. These two-scale methods have a long history in mathematics \cite{bensoussan2011asymptotic, allaire1992homogenization} and have been applied to a variety of settings including optics \cite{touboul2023high, craster2011high}, phononic crystals \cite{ammari2019bloch}, semiconductors \cite{janssen2016precise}, elastic composites \cite{boutin2014large}, frame and lattice structures \cite{colquitt2015high, nolde2011high}, materials with imperfect interfaces \cite{assier2020high, guzina2021effective} and forced problems \cite{guzina2019rational}. Most relevant to our work are methods for discrete systems \cite{craster2010lattice} and for describing localization due to defects \cite{makwana2013localised, craster2023asymptotic}.

The two-scale effective model proposed in this work characterizes eigenvectors near to the localization transition as the product of a short-scale modified standing mode and a long-scale amplitude function. The short-scale modified standing mode is the eigenvector that exists when the system is exactly at the localization transition point. This is the non-Hermitian analogue of the reference Bloch mode used in $\mathbf{k}\cdot\mathbf{p}$ theory and HFH \cite{voonwillatzen2009kp, janssen2016precise, craster2010high, craster2023asymptotic}. We derive an equation for the long-scale amplitude function which characterizes the effective properties of the system in an asymptotic neighbourhood of the transition point. Solutions of this effective model will be either exponentially decaying or growing away from the defect, revealing whether an eigenvector is either localized at the defect or the edge, respectively. The switch between these two regimes, captured by the effective model, is the localization transition in question.

To demonstrate our two-scale method, we consider a model that is a variant of the non-Hermitian Hamiltonians introduced by Hatano and Nelson \cite{hatano1996localization, hatano1998nonHermitian}, which has non-reciprocal coupling terms and a modulated on-site potential function. In order to illuminate the fundamental physics of the localization transition, we consider the simplest model that captures its key features. This is a homogeneous potential with a single defect in the center of the array. We derive the critical defect size at which the localization transition occurs and introduce a two-scale ansatz to derive an effective model that holds close to this point. This gives quantitative predictions for eigenvectors on either side of the transition, which are localized either at the edge or at the defect, as captured by the effective model.




\begin{figure}
    \centering
    \includegraphics[width=\linewidth]{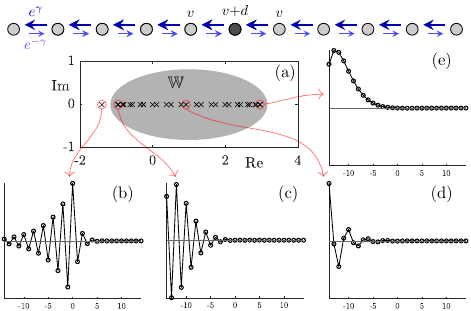}
    \caption{Hamiltonians with non-reciprocal couplings exhibit the non-Hermitian skin effect whereby eigenvalues falling within a specified region $\mathbb{W}$ are all localized at one edge of the system, as shown in (c), (d) and (e). All the eigenvalues for this 29-site system are shown in the complex plane in (a), along with the region $\mathbb{W}$. When a defect $d$ is added to the potential at one of the sites, this can cause an eigenvalue to leave $\mathbb{W}$ and be localized at the defect site, as shown in (b). We propose a two-scale effective model for this transition.}
    \label{fig:1}
\end{figure}

\textit{The unperturbed skin effect}.---We are interested in the eigenvalues of the non-Hermitian Hamiltonian
\begin{equation} \label{eq:Hamiltonian}
    H\psi_n=V_n\psi_n+e^{-\gamma}\psi_{n-1}+e^{\gamma}\psi_{n+1},
\end{equation}
where $V_n:\mathbb{Z}\to\mathbb{R}$ is some potential function and $\gamma>0$ is a real constant that characterizes the non-reciprocity of the system.

In the unperturbed case, when $V_n=v\in\mathbb{R}$ for all $n$, \eqref{eq:Hamiltonian} is a truncated Toeplitz matrix with symbol given by
\begin{equation}
    f_H(s)=v-e^\gamma s - e^{-\gamma}s^{-1},
\end{equation}
for $s\in S^1\subset\mathbb{C}$, the unit circle in the complex plane \cite{bottcher2012introduction}. Let $\mathbb{W}$ be the set of all $z\in\mathbb{C}$ such that the winding number of $f_H$ around $z$ is negative. Then, it is a general property of truncated Toeplitz matrices that any eigenvalue falling within $\mathbb{W}$ will correspond to an eigenvector that is exponentially localized at one edge \cite{bottcher2012introduction, trefethen2005Spectra}. This phenomenon is the non-Hermitian skin effect. Some examples of eigenvectors that are localized at an edge due to the skin effect are shown in Fig.~\ref{fig:1}(c)-(e). The corresponding eigenvalues all fall within the shaded grey region $\mathbb{W}$ in the complex plane in Fig.~\ref{fig:1}(a). This spectrum is that of a system of 29 sites with a defect at the center. This defect is sufficiently large that one of the eigenvalues has escaped $\mathbb{W}$ and the corresponding eigenvector is localized at the defect site, as shown in Fig.~\ref{fig:1}(b). Since the eigenvalues are always real valued, they can only leave the region $\mathbb{W}$ at the points $E=v\pm2\cosh\gamma$. How the eigenvectors change as the eigenvalues cross these points is the transition we will capture in this work.

\begin{figure*}
    \centering
    \includegraphics[width=\linewidth]{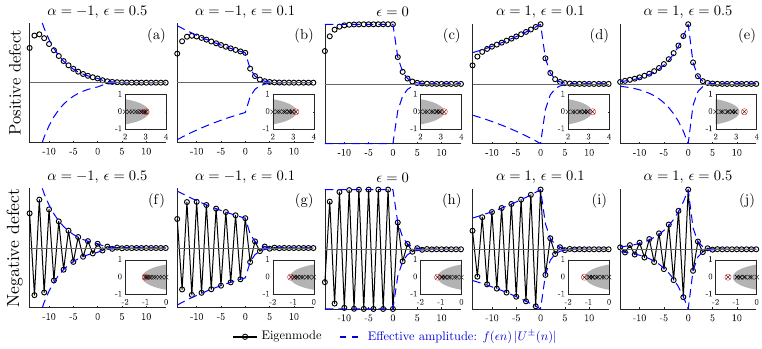}
    \caption{The maximal or minimal eigenvectors of a 29-site system with a defect in the center. (a)-(e) show the eigenvector with the largest eigenvalue in a system with a positive defect $d=2\sinh\gamma+\alpha\epsilon$. The effective amplitude predicted by our two-scale model \eqref{eq:effective} is shown in a dotted line. Inset shows the eigenvalues in the complex plane with $\mathbb{W}$ shaded and the eigenvalue of the plotted eigenvector circled in red. (f)-(j) show the analogous eigenpairs for a negative defect $d=-(2\sinh\gamma+\alpha\epsilon)$. $\gamma=0.4$ and $v=1$ are used throughout and the values of $\alpha\in\{-1,1\}$ and $\epsilon>0$ are shown above each plot.}
    \label{fig:2}
\end{figure*}

\textit{Defect-induced localization}.---To examine the fundamental mechanism of the transition from localization due to the skin effect to defect-induced localization in the bulk, we study the spectrum of the Hamiltonian \eqref{eq:Hamiltonian} with a single defect in the on-site potential. That is, we suppose that $V_n=v+ d \delta_{n,0}$ where $d\in\mathbb{R}$ characterizes the magnitude of the defect and $\delta_{n,m}$ is the Kronecker delta. Adding a small defect will not be sufficient to cause localization in the bulk as the eigenvalues will still be within $\mathbb{W}$. The localization transition occurs at a non-zero defect size given by $d=\pm2\sinh\gamma$. This critical defect size causes an eigenvalue to take the value $E=v\pm2\cosh\gamma$ which falls at the transition points on the boundary of $\mathbb{W}$. The critical defect size does not depend on the background potential $v$ since changing $v$ translates both the spectrum and the region $\mathbb{W}$. The corresponding eigenvectors have the form
\begin{equation} \label{eq:Upm}
    U^{\pm}(n)=\begin{cases}
        (\pm 1)^n & n\leq 0,\\
        (\pm e^{-2\gamma})^n & n>0.
    \end{cases}
\end{equation}
They are modified standing modes in the sense that they are standing modes (which are either periodic or anti-periodic) on the left of the defect and decay quickly to the right. Variants of $U^{\pm}$ for a finite-sized system are shown in Figs.~\ref{fig:2}(c) and \ref{fig:2}(h).

To develop an effective model for the transition we consider a system that is an $\epsilon$-perturbation of the critical system for some $0<\epsilon\ll1$. We consider the potential
\begin{equation}
    V_n=v\pm(2\sinh\gamma + \alpha\epsilon) \delta_{n,0},
\end{equation}
where $\alpha\in\{-1,1\}$ captures the sign of the perturbation away from the critical defect size. 

The eigenvectors plotted in Figs.~\ref{fig:1} and~\ref{fig:2} display typical two-scale behaviour: oscillations on a short length scale with a slowly varying amplitude that is modulated on a longer scale. To capture this, we introduce the continuous long-scale variable $\eta=\epsilon n$ and seek a two-scale solution as an asymptotic series $\psi_n=\Psi^{(0)}(\eta,n)+\epsilon\Psi^{(1)}(\eta,n)+\dots$ with eigenvalue $E=E^{(0)}+\epsilon E^{(1)}+\epsilon^2 E^{(2)}+\dots$. Details of this asymptotic analysis can be found in the Supplemental Material, which has some crucial differences from the traditional application to Hermitian systems. The leading-order form of the eigenvector is 
\begin{equation} \label{eq:Psi0}
    \Psi^{(0)}(\eta,n)=f(\eta)U^{\pm}(n),
\end{equation}
where $U^{\pm}(n)$ are the modified standing modes given in \eqref{eq:Upm} and $f(\eta)$ is a long-scale amplitude modulation that solves the Schr\"odinger eigenvalue problem
\begin{equation} \label{eq:effective}
    \cosh\gamma f_{\eta\eta}(\eta)+\alpha \delta(\eta)f(\eta)=\pm E^{(2)} f(\eta).
\end{equation}
This effective model characterizes the transition between localization at the edge and at the defect site. It has a solution which decays away from the defect (so that $f\to0$ as $\eta\to\pm\infty$) if and only if $\alpha>0$. Conversely, when $\alpha<0$ any solution of \eqref{eq:effective} must be exponentially growing away from the defect, meaning the eigenvector will be localized at the edge. Thus, the model shows that there can be localization at the defect site if and only if the defect is larger than the critical size. The first eigensolution of \eqref{eq:effective} is
\begin{equation} \label{eq:fsoln}
    f(\eta)=\exp\left( -\frac{\alpha}{2\cosh\gamma}|\eta| \right),
\end{equation}
with eigenvalue $E^{(2)}=\pm (4\cosh\gamma)^{-1}$. Collecting the other terms in the asymptotic expansion for the eigenvalue (see the Supplemental Material for details) gives
\begin{equation} \label{eq:Eval}
    E=v \pm 2\cosh\gamma \pm \epsilon\alpha\tanh\gamma \pm \epsilon^2 \frac{1}{4\cosh\gamma} +O(\epsilon^3).
\end{equation}

The leading-order approximation of the eigenvector \eqref{eq:Psi0} is compared to numerical solutions in Fig.~\ref{fig:2}. The numerical solutions are obtained for a finite-sized system with 29 sites and Dirichlet boundary conditions at either ends. As a result, the eigenvectors experience some edge effects (visible at the left edge) which are not accounted for in the effective model. However, we see very good agreement away from the edges including, crucially, near to the defect site. First, we observe in \ref{fig:2}(c) and \ref{fig:2}(h) that when $\epsilon=0$ the eigenvector is the modified standing mode given by $U^\pm$ in \eqref{eq:Upm}. When $\alpha<0$ the eigenvectors are localized at the left edge of the system due to the skin effect. This is shown in Figs.~\ref{fig:2}(a), \ref{fig:2}(b), \ref{fig:2}(f) and \ref{fig:2}(g), where we see that the effective amplitude (with $f(\eta)$ being exponentially growing away from the defect) is accurately predicting the profile of the eigenvector near to the defect. Finally, when $\alpha>0$ the modes are localized at the defect site ($n=0$). Again, the effective amplitude gives a good prediction of the profile (with the localization being due to the fact that $f(\eta)$ is exponentially decaying in this case). The good agreement shown in Fig.~\ref{fig:2} is in spite of the fact that the system is relatively small (and the eigenvectors experience edge effects at the ends of the array) and holds even for relatively large values of the asymptotic parameter $\epsilon$.



\begin{figure}
    \centering
    \includegraphics[width=\linewidth]{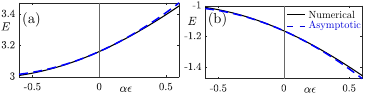}
    \caption{The maximal or minimal eigenvalues for a system with either (a) a positive defect or (b) a negative defect, respectively. Numerical values for a 29-site system with a single defect $d=\pm(2\sinh\gamma+\alpha\epsilon)$ at the center are compared to the asymptotic formula \eqref{eq:Eval} for $\gamma=0.4$ and $v=1$.}
    \label{fig:3}
\end{figure}

The asymptotic formula \eqref{eq:Eval} for the minimal or maximal eigenvalue (which may escape $\mathbb{W}$ for sufficiently large defects) is compared to numerical values in Fig.~\ref{fig:3}. The same system of 29 sites from Fig.~\ref{fig:2} is used. Again, we see excellent agreement, even for large $\epsilon$ and in spite of the edge effects that occur in the finite-sized system.

\textit{Conclusions}.--- Our results show that, despite the exotic features of non-Hermitian systems and their sometimes counter-intuitive localization properties, defect-induced localization transitions can be described concisely and intuitively. Our effective model means that the competition between localization at the edge and at a defect site within the bulk can be predicted without intensive computations. As non-Hermitian systems are increasingly used in applications such as sensing \cite{budich2020sensors, franca2022non}, tunneling \cite{jana2023tunneling} and more \cite{bergholtz2021exceptional, zhang2022review}, it is imperative to have at hand a broad array of analytical and numerical tools. Our results add a valuable new approach to the toolbox for studying localization phenomena in non-Hermitian systems.

The two-scale effective model developed here gives insight into the fundamental mechanism that causes eigenvectors to transition between localization at an edge and at a defect site in the bulk due. This effective model reduces localization at either the edge or the defect site to the growth or decay of a long-scale modulation function. Meanwhile, the short-scale oscillations of eigenvectors are captured by a modified standing wave. As well as providing new intuition, this model gives accurate predictions of the profiles of eigenvectors on either side of the localization transition, even for large values of the asymptotic parameter and in small arrays (with edge effects due to fixed boundary conditions). 

An exciting future direction for investigation is extending these methods to non-reciprocal lattices with potentials given by independent random variables on multiple sites (known variously as the non-Hermitian Anderson model \cite{hatano1996localization, hatano1998nonHermitian, brouwer1997theory, goldsheid2018real}). Our work considered the simplest possible model that displayed the essential features of the transition between skin and defect localization. This has illuminated the fundamental mechanism responsible for the competition between these two localization effects.

\begin{acknowledgments}
B.D. is supported by EPSRC under grant number EP/X027422/1. R.V.C.'s work was funded by UK Research and Innovation (UKRI) under the UK government’s Horizon Europe funding guarantee [grant number 10033143].
\end{acknowledgments}

\bibliography{references}

\clearpage
\onecolumngrid

\begin{center}
    \textbf{\large Supplementary Material}
\end{center}

\section{Positive defect}
We are interested in solutions of the difference equation
\begin{equation}
    v\psi_n + (2\sinh\gamma+\alpha\epsilon)\delta_{n,0}\psi_n+e^{-\gamma}\psi_{n-1}+e^{\gamma}\psi_{n+1}=E\psi_n,
\end{equation}
where the defect $d=2\sinh\gamma+\alpha\epsilon$ has been chosen to be an $O(\epsilon)$ perturbation away from the critical defect size $d=2\sinh\gamma$ (which leads to the largest eigenvalue being exactly on the boundary of $\mathbb{W}$). We introduce the continuous long-scale variable $\eta=\epsilon n$ and will seek a two-scale solution in terms of $n$ and $\eta$. For convenience, we elect to switch the $O(\epsilon)$ part of the defect onto the long-scale variable $\eta$. To do so, we make the association $\delta_{n,0}=\epsilon\delta(\eta)$. This is a standard equivalence (used in \emph{e.g.} \cite{makwana2013localised}) and is taken to give the appropriate normalization between the Kronecker and Dirac deltas when re-scaling. This can be seen formally by comparing the expressions
\begin{equation}
    1=\int_{-\infty}^{\infty} \delta(\eta) \ \mathrm{d}\eta =\int_{-\infty}^{\infty} \delta(\epsilon n)\epsilon \ \mathrm{d}n,
\end{equation}
and
\begin{equation}
    1=\sum_{n=-\infty}^{\infty} \delta_{n,0} =\int_{-\infty}^{\infty} \delta_{n,0} \ \mathrm{d}n,
\end{equation}
where integration with respect to the discrete variable $n$ should use the counting measure. This leads to the difference equation
\begin{equation} \label{eq:diffpos}
    v\psi_n + 2\sinh\gamma\delta_{n,0}\psi_n+\alpha\epsilon^2\delta(\eta)\psi_n+e^{-\gamma}\psi_{n-1}+e^{\gamma}\psi_{n+1}=E\psi_n.
\end{equation}

We seek a two-scale solution $\psi_n=\Psi(\eta,n)$. We can Taylor expand in the long-scale variable to see that
\begin{align}
    \psi_{n-1}&= \Psi(\eta-\epsilon,n-1) = \Psi(\eta,n-1) - \epsilon \Psi_\eta(\eta,n-1) +\frac{1}{2}\epsilon^2 \Psi_{\eta\eta}(\eta,n-1) +\dots, \label{eq:psi-1asym}\\
    \psi_{n+1}&= \Psi(\eta+\epsilon,n+1) = \Psi(\eta,n+1) + \epsilon \Psi_\eta(\eta,n+1) +\frac{1}{2}\epsilon^2 \Psi_{\eta\eta}(\eta,n+1) +\dots, \label{eq:psi1asym}
\end{align}
where the subscripts are used to denote differentiation with respect to that variable. In addition to this Taylor expansion, we will seek solutions of \eqref{eq:diffpos} in terms of asymptotic expansions:
\begin{align}
    \Psi&=\Psi^{(0)}+\epsilon\Psi^{(1)}+\epsilon^2\Psi^{(2)}+\dots, \label{eq:Psiasym} \\
    E&=E^{(0)}+\epsilon E^{(1)}+\epsilon^2 E^{(2)}+\dots. \label{eq:Easym}
\end{align}

Substituting \eqref{eq:psi-1asym}--\eqref{eq:Easym} into \eqref{eq:diffpos} yields a hierarchy of equations in $\epsilon$. At leading order, we have
\begin{equation}
    v\Psi^{(0)}(\eta,n)+2\sinh\gamma\delta_{n,0}\Psi^{(0)}(\eta,n)+e^{-\gamma}\Psi^{(0)}(\eta,n-1)+e^{\gamma}\Psi^{(0)}(\eta,n+1)=E^{(0)}\Psi^{(0)}(\eta,n).
\end{equation}
This has solution
\begin{equation} \label{eq:Psi0}
    \Psi^{(0)}(\eta,n)=f(\eta)U^{+}(n),
\end{equation}
with eigenvalue $E^{(0)}=v+2\cosh\gamma$, where 
\begin{equation} \label{eq:U0}
    U^{+}(n)=\begin{cases}
        1 & n\leq 0,\\
        (e^{-2\gamma})^n & n>0.
    \end{cases}
\end{equation}
and $f(\eta)$ is some long-scale function that is not yet determined. This form of the leading-order solution \eqref{eq:Psi0} is consistent with the traditional applications of high-frequency homogenization to Hermitian systems \cite{craster2010high}. It is given by $U^{+}(n)$, which is the mode that exists when $\epsilon=0$ (and is reminiscent of a standing wave, at least on the left of the defect site), modulated by some slowly-varying amplitude function $f(\eta)$. Determining $f(\eta)$, which describes the long-scale variation of the eigenmode at leading order and thus captures whether the eigenmode is localized at the defect or not, is the primary goals of this analysis.

Collecting the terms at order $\epsilon$, we have the equation
\begin{equation}
    \begin{split}
    v\Psi^{(1)}(\eta,n)+2\sinh\gamma\delta_{n,0}\Psi^{(1)}(\eta,n)+e^{-\gamma}\Psi^{(1)}(\eta,n-1)-e^{-\gamma}\Psi^{(0)}_\eta(\eta,n-1) &+e^{\gamma}\Psi^{(1)}(\eta,n+1)+e^{\gamma}\Psi^{(0)}_\eta(\eta,n+1)\\
    &=E^0\Psi^{(1)}(\eta,n)+E^{(1)}\Psi^{(0)}(\eta,n).
    \end{split}
\end{equation}
A solution for $\Psi^{(1)}$ with the same form as \eqref{eq:Psi0} exists provided that 
\begin{equation}
    -e^{-\gamma}\Psi^{(0)}_\eta(\eta,n-1)+e^{\gamma}\Psi^{(0)}_\eta(\eta,n+1)=E^{(1)}\Psi^{(0)}(\eta,n).
\end{equation}
Inserting the expression \eqref{eq:Psi0} for $\Psi^{(0)}$ gives different behaviour depending on the sign of $n$ (and, consequently, of $\eta$). If $n<0$ we have that 
\begin{equation}
    2\sinh\gamma f_\eta(\eta)=E^{(1)} f(\eta),
\end{equation}
whereas for $n>0$ 
\begin{equation}
    -2\sinh\gamma f_\eta(\eta)=E^{(1)} f(\eta).
\end{equation}
We elect to capture this sign change using the continuous variable $\eta$ (since $\mathrm{sgn}(\eta)=\mathrm{sgn}(n)$, where $\mathrm{sgn}(x)=|x|/x$ is the standard sign function), yielding the first-order ODE
\begin{equation} \label{eq:ODE1f}
    2\sinh\gamma f_\eta(\eta)=-E^{(1)} \mathrm{sgn}(\eta) f(\eta).
\end{equation}
This equation represents a crucial difference from the application of high-frequency homogenization to the corresponding Hermitian system, as in \cite{makwana2013localised}. When $\gamma=0$ and the system is Hermitian, the left-hand side vanishes meaning $E^{(1)}=0$ and the defect eigenvalue is an $O(\epsilon^2)$ perturbation of $E^{(0)}$. In this case, if $\gamma\neq0$ then $E^{(1)}\neq 0$. However, \eqref{eq:ODE1f} is not sufficient to determine either the function $f(\eta)$ or the constant $E^{(1)}$. We must proceed to higher orders of $\epsilon$ and return to \eqref{eq:ODE1f} later.

We can also collect the terms at order $\epsilon^2$, which yields the equation 
\begin{equation}
    \begin{split}
    & v\Psi^{(2)}(\eta,n)+2\sinh\gamma\delta_{n,0}\Psi^{(2)}(\eta,n)+\alpha\delta(\eta)\Psi^{(0)}(\eta,n)+e^{-\gamma}\Psi^{(2)}(\eta,n-1)-e^{-\gamma}\Psi^{(1)}_\eta(\eta,n-1) +\frac{1}{2} e^{-\gamma}\Psi^{(0)}_{\eta\eta}(\eta,n-1)
    \\&\quad +e^{\gamma}\Psi^{(2)}(\eta,n+1)+e^{\gamma}\Psi^{(1)}_\eta(\eta,n+1)+\frac{1}{2} e^{\gamma}\Psi^{(0)}_{\eta\eta}(\eta,n+1)=E^{(0)}\Psi^{(2)}(\eta,n)+E^{(1)}\Psi^{(1)}(\eta,n)+E^{(2)}\Psi^{(0)}(\eta,n).
    \end{split}
\end{equation}
Once again, a solution for $\Psi^{(2)}$ with short-scale dependence given by \eqref{eq:U0} can be constructed provided that the other terms cancel. Assuming the $\eta$-dependent part of $\Psi^{(1)}$ also satisfies \eqref{eq:ODE1f}, we are left with
\begin{equation}
    \alpha\delta(\eta)\Psi^{(0)}(\eta,n)+\frac{1}{2} e^{-\gamma}\Psi^{(0)}_{\eta\eta}(\eta,n-1)+\frac{1}{2} e^{\gamma}\Psi^{(0)}_{\eta\eta}(\eta,n+1)=E^{(2)}\Psi^{(0)}(\eta,n).
\end{equation}
Substituting $\Psi^{(0)}(\eta,n)=f(\eta)U^{+}(n)$ from \eqref{eq:Psi0} gives
\begin{equation}
    \alpha\delta(\eta)f(\eta)U^{+}(n)+\frac{1}{2} e^{-\gamma}f_{\eta\eta}(\eta)U^{+}(n)+\frac{1}{2} e^{\gamma}f_{\eta\eta}(\eta)U^{+}(n)=E^{(2)}f(\eta)U^{+}(n),
\end{equation}
for $n<0$ and
\begin{equation}
    \alpha\delta(\eta)f(\eta)U^{+}(n)+\frac{1}{2} e^{-\gamma}e^{2\gamma}f_{\eta\eta}(\eta)U^{+}(n)+\frac{1}{2} e^{\gamma}e^{-2\gamma}f_{\eta\eta}(\eta)U^{+}(n)=E^{(2)}f(\eta)U^{+}(n),
\end{equation}
for $n>0$. In either case, this can be rearranged to give a Schr\"odinger eigenvalue problem $f(\eta)$ and $E^{(2)}$:
\begin{equation} \label{eq:effectivepos}
    \cosh\gamma f_{\eta\eta}(\eta)+\alpha \delta(\eta)f(\eta)=E^{(2)} f(\eta).
\end{equation}

The Schr\"odinger eigenvalue problem \eqref{eq:effectivepos} is our effective model for the localization at the defect site. We observe, firstly, that it has a decaying solution $f(\eta)\to0$ as $\eta\to\pm\infty$ for positive $E^{(2)}$ if and only if $\alpha>0$. To see this, we multiply \eqref{eq:effectivepos} by $f$ and integrate to obtain
\begin{equation}
    \left[\cosh\gamma f_\eta f\right]_\infty^\infty-\cosh\gamma\int_{-\infty}^\infty (f_\eta)^2  \ \mathrm{d}\eta +\alpha f(0)^2 = E^{(2)} \int_{-\infty}^\infty f^2 \ \mathrm{d}\eta,
\end{equation}
from which it is clear that $\lim_{\eta\to\pm\infty}f_\eta f=0$ only if $\alpha>0$ (since we expect $E^{(2)}>0$ so that the eigenvalue will fall outside of $\mathbb{W}$ and not be localized at the left edge of the system due to the skin effect). When $\alpha>0$, the solution can be found using Fourier transforms. We let $\alpha=1$ and then in Fourier space \eqref{eq:effectivepos} becomes
\begin{equation}
    -\cosh\gamma\xi^2 \hat f(\xi)+  f(0)=E^{(2)} \hat f(\xi),
\end{equation}
from which we find that 
\begin{equation}
    \hat f(\xi) = \frac{ f(0)}{E^{(2)}+\cosh\gamma \xi^2}.
\end{equation}
Applying the inverse Fourier transform yields the solution
\begin{equation}
    f(\eta)=\frac{f(0)}{2\sqrt{E^{(2)}\cosh\gamma}} \exp\left( -\sqrt{\frac{E^{(2)}}{\cosh\gamma}}|\eta| \right).
\end{equation}
Inspecting the value at $\eta=0$ shows that $2\sqrt{E^{(2)}\cosh\gamma}=1$, from which we have that
\begin{equation}
    E^{(2)}=\frac{1}{4\cosh\gamma}.
\end{equation}
Fixing the normalization of $f(\eta)$ to be such that $\sup_\eta |f(\eta)|=f(0)=1$, gives that
\begin{equation} \label{eq:fsolnpos}
    f(\eta)=\exp\left( -\frac{1}{2\cosh\gamma}|\eta| \right).
\end{equation}
Now that we have identified $f(\eta)$, we can use \eqref{eq:ODE1f} to find the value of $E^{(1)}$. Differentiating \eqref{eq:fsolnpos} yields
\begin{equation}
    f_\eta(\eta)=-\frac{1}{2\cosh\gamma} \mathrm{sgn}(\eta) f(\eta),
\end{equation}
which can be compared with \eqref{eq:ODE1f} to see that
\begin{equation}
    E^{(1)}=\tanh\gamma.
\end{equation}

Bringing this all together, we have found that, if and only if $\alpha=1>0$, there exists an eigenmode localized at the defect site $n=0$ which has eigenvalue
\begin{equation}
    E=v+2\cosh\gamma + \epsilon \tanh\gamma + \epsilon^2 \frac{1}{4\cosh\gamma} +O(\epsilon^3),
\end{equation}
and leading-order profile 
\begin{equation}
    \psi_n= U^{+}(n) \exp\left( -\frac{1}{2\cosh\gamma}|\epsilon n| \right) + O(\epsilon),
\end{equation}
where $U^{+}(n)$ was specified in \eqref{eq:U0}.

When $\alpha<0$, the effective model \eqref{eq:effectivepos} is still solvable but its solution will not decay as $\eta\to\pm\infty$. Indeed, we expect the general solution of \eqref{eq:effectivepos} to be of the form $f(\eta)=A\exp(-\sqrt{E^{(2)}/\cosh\gamma}\eta)+B\exp(\sqrt{E^{(2)}/\cosh\gamma}\eta)$. The constants $A$ and $B$ will be different for $\eta<0$ and $\eta>0$ to account for the discontinuity in the derivative $f_\eta(\eta)$ at $\eta=0$ that is introduced by the $\delta(\eta)$ term. Matching the solutions such that $f(\eta)$ is continuous and $f_\eta(\eta)$ has a jump of magnitude $(\cosh\gamma)^{-1}$ at $\eta=0$ leads to the normalized solution
\begin{equation} 
    f(\eta)=\exp\left( \frac{1}{2\cosh\gamma}|\eta| \right),
\end{equation}
where we have let $\alpha=-1$. Noting that
\begin{equation} \label{eq:firstderiv}
    f_\eta(\eta)= \mathrm{sgn}(\eta) \frac{1}{2\cosh\gamma} f(\eta),
\end{equation}
and 
\begin{equation}
    f_{\eta\eta}(\eta)= \frac{1}{4(\cosh\gamma)^2} f(\eta) + \frac{1}{\cosh\gamma}\delta(\eta) f(\eta),
\end{equation}
we can see that substitution into \eqref{eq:effectivepos} yields the corresponding eigenvalue $E^{(2)}=(4\cosh\gamma)^{-1}$ (which is unchanged from $\alpha<0$). However, the switch from decay to growth on either side of the defect has led to a change of sign in \eqref{eq:firstderiv}, such that substitution into \eqref{eq:ODE1f} leads to $E^{(1)}=-\tanh\gamma$. 

Combining the analysis for the $\alpha>0$ and $\alpha<0$ cases, we find the leading-order profile of the eigenvector to be
\begin{equation}
    \psi_n= U^{+}(n) \exp\left( -\frac{\alpha}{2\cosh\gamma}|\epsilon n| \right) + O(\epsilon),
\end{equation}
with associated eigenvalue
\begin{equation}
    E=v+2\cosh\gamma + \epsilon\alpha \tanh\gamma + \epsilon^2 \frac{1}{4\cosh\gamma} +O(\epsilon^3).
\end{equation}

\section{Negative defect}

The above analysis can be repeated for the case of a negative defect. Here, we study the difference equation
\begin{equation} \label{eq:diffneg}
    v\psi_n - 2\sinh\gamma\delta_{n,0}\psi_n-\alpha\epsilon^2\delta(\eta)\psi_n+e^{-\gamma}\psi_{n-1}+e^{\gamma}\psi_{n+1}=E\psi_n,
\end{equation}
and proceed to seek solutions in terms of similar expansions similar to \eqref{eq:Psiasym} and \eqref{eq:Easym}. The leading-order equation is 
\begin{equation}
    v\Psi^{(0)}(\eta,n)-2\sinh\gamma\delta_{n,0}\Psi^{(0)}(\eta,n)+e^{-\gamma}\Psi^{(0)}(\eta,n-1)+e^{\gamma}\Psi^{(0)}(\eta,n+1)=E^{(0)}\Psi^{(0)}(\eta,n).
\end{equation}
This admits the solution
\begin{equation}
    \Psi^{(0)}(\eta,n)=f(\eta)U^-(n),
\end{equation}
with eigenvalue $E^0=v-2\cosh\gamma$, where 
\begin{equation} \label{eq:U0neg}
    U^-(n)=\begin{cases}
        (-1)^n & n\leq 0,\\
        (-e^{-2\gamma})^n & n>0.
    \end{cases}
\end{equation}
Once again, the goal is to find the long-scale function $f(\eta)$.

Collecting the terms at order $\epsilon$ gives the equation
\begin{equation}
    \begin{split}
    v\Psi^{(1)}(\eta,n)-2\sinh\gamma\delta_{n,0}\Psi^{(1)}(\eta,n) &+e^{-\gamma}\Psi^{(1)}(\eta,n-1)-e^{-\gamma}\Psi^{(0)}_\eta(\eta,n-1) +e^{\gamma}\Psi^{(1)}(\eta,n+1)\\
    &\quad+e^{\gamma}\Psi^{(0)}_\eta(\eta,n+1)=E^{(0)}\Psi^{(1)}(\eta,n)+E^{(1)}\Psi^{(0)}(\eta,n).
    \end{split}
\end{equation}
As above, we want that
\begin{equation}
    -e^{-\gamma}\Psi^{(0)}_\eta(\eta,n-1)+e^{\gamma}\Psi^{(0)}_\eta(\eta,n+1)=E^{(1)}\Psi^{(0)}(\eta,n),
\end{equation}
which leads to the first-order ODE
\begin{equation} \label{eq:ODE1fneg}
    f_\eta(\eta)=\frac{E^{(1)}}{2\sinh\gamma} \mathrm{sgn}(\eta) f(\eta),
\end{equation}
which is the same as \eqref{eq:ODE1f} up to a change of sign of the constant on the right-hand side.

Proceeding to the $\epsilon^2$ terms, we find the equation
\begin{equation}
    \begin{split}
    v\Psi^{(2)}(\eta,n)&-2\sinh\gamma\delta_{n,0}\Psi^{(2)}(\eta,n)-\alpha\delta(\eta)\Psi^0(\eta,n)+e^{-\gamma}\Psi^2(\eta,n-1)-e^{-\gamma}\Psi^1_\eta(\eta,n-1) +\frac{1}{2} e^{-\gamma}\Psi^0_{\eta\eta}(\eta,n-1)
    \\&\quad+e^{\gamma}\Psi^2(\eta,n+1)+e^{\gamma}\Psi^1_\eta(\eta,n+1)+\frac{1}{2} e^{\gamma}\Psi^0_{\eta\eta}(\eta,n+1)=E^0\Psi^2(\eta,n)+E^1\Psi^1(\eta,n)+E^2\Psi^0(\eta,n).
    \end{split}
\end{equation}
Using the same logic as above, we arrive at the equation
\begin{equation}
    -\alpha\delta(\eta)f(\eta)U^-(n)-\frac{1}{2} e^{-\gamma}f_{\eta\eta}(\eta)U^-(n-1)-\frac{1}{2} e^{\gamma}f_{\eta\eta}(\eta)U^-(n+1)=E^{(2)}f(\eta)U^-(n),
\end{equation}
which leads to a Schr\"odinger eigenvalue problem for $f(\eta)$ and $E^{(2)}$, which is analogous to \eqref{eq:effectivepos}:
\begin{equation} \label{eq:effectiveneg}
    \cosh\gamma f_{\eta\eta}(\eta)+\alpha \delta(\eta)f(\eta)=-E^{(2)} f(\eta).
\end{equation}
Repeating the analysis of \eqref{eq:effectivepos} but with the sign of the eigenvalue changed, we see that \eqref{eq:effectiveneg} admits a decaying solution for $f(\eta)$ if and only if $\alpha>0$ and $E^{(2)}<0$. Thus, letting $\alpha=1$ leads to the solution
\begin{equation} \label{eq:fdecaying}
    f(\eta)=\exp\left( -\frac{1}{2\cosh\gamma}|\eta| \right),
\end{equation}
along with
\begin{equation}
    E^{(2)}=-\frac{1}{4\cosh\gamma}.
\end{equation}
Using \eqref{eq:ODE1fneg}, this leads to 
\begin{equation}
    E^{(1)}=-\tanh\gamma.
\end{equation}

Finally, we can conclude that, if and only if $\alpha=1>0$, there exists an eigenmode localized at the defect site $n=0$ which has eigenvalue
\begin{equation}
    E=v-2\cosh\gamma - \epsilon \tanh\gamma - \epsilon^2 \frac{1}{4\cosh\gamma} +O(\epsilon^3),
\end{equation}
and leading-order profile 
\begin{equation}
    \psi_n= U^{-}(n) \exp\left( -\frac{1}{2\cosh\gamma}|\epsilon n| \right) + O(\epsilon),
\end{equation}
where $U^{-}(n)$ was specified in \eqref{eq:U0neg}.

The analysis for the $\alpha<0$ case is, again, similar to the previous section. We find that the exponentially decaying solution \eqref{eq:fdecaying} switches to an exponentially growing mode. Thus, we obtain the general expressions for the eigenvector
\begin{equation}
    \psi_n= U^{-}(n) \exp\left( -\frac{\alpha}{2\cosh\gamma}|\epsilon n| \right) + O(\epsilon),
\end{equation}
and its associated eigenvalue
\begin{equation}
    E=v-2\cosh\gamma - \epsilon\alpha \tanh\gamma - \epsilon^2 \frac{1}{4\cosh\gamma} +O(\epsilon^3),
\end{equation}
which hold for $\alpha\in\{-1,1\}$.

\end{document}